\documentclass[prl,preprint,amsfonts,superscriptaddress]{revtex4-1}
\usepackage{dcolumn}
\usepackage{amsmath}
\usepackage{amssymb}
\usepackage{graphicx}
\usepackage{bm}

\begin{document}

\graphicspath{ {figures/} }

\title{Probing the electron-phonon interaction in correlated systems with coherent lattice fluctuation spectroscopy}

\vspace{2cm}

\author{Andreas Mann}
	\affiliation{Laboratory for Ultrafast Microscopy and Electron Scattering, ICMP, Ecole Polytechnique F\'{e}d\'{e}rale de Lausanne, CH-1015 Lausanne, Switzerland}

\author{Edoardo Baldini}
	\affiliation{Laboratory for Ultrafast Microscopy and Electron Scattering, ICMP, Ecole Polytechnique F\'{e}d\'{e}rale de Lausanne, CH-1015 Lausanne, Switzerland}
	\affiliation{Laboratory of Ultrafast Spectroscopy, ISIC, Ecole Polytechnique F\'{e}d\'{e}rale de Lausanne, CH-1015 Lausanne, Switzerland}

\author{Antonio Tramontana}
	\affiliation{Institute for Complex Systems - Consiglio Nazionale delle Ricerche, and Physics Department, University of Rome La Sapienza, I-00185 Rome, Italy}

\author{Ekaterina Pomjakushina}
	\affiliation{Solid State Chem. Group, Laboratory for Developments and Methods, Paul Scherrer Institute, CH-5232 Villigen PSI, Switzerland}

\author{Kazimierz Conder}
	\affiliation{Solid State Chem. Group, Laboratory for Developments and Methods, Paul Scherrer Institute, CH-5232 Villigen PSI, Switzerland}

\author{Christopher Arrell}
	\affiliation{Laboratory of Ultrafast Spectroscopy, ISIC, Ecole Polytechnique F\'{e}d\'{e}rale de Lausanne, CH-1015 Lausanne, Switzerland}
	
\author{Frank van Mourik}
	\affiliation{Laboratory of Ultrafast Spectroscopy, ISIC, Ecole Polytechnique F\'{e}d\'{e}rale de Lausanne, CH-1015 Lausanne, Switzerland}

\author{Jos{\'e} Lorenzana}
	\affiliation{Institute for Complex Systems - Consiglio Nazionale delle Ricerche, and Physics Department, University of Rome La Sapienza, I-00185 Rome, Italy}

\author{Fabrizio Carbone}
	\affiliation{Laboratory for Ultrafast Microscopy and Electron Scattering, ICMP, Ecole Polytechnique F\'{e}d\'{e}rale de Lausanne, CH-1015 Lausanne, Switzerland}

\date{\today}

\begin{abstract}
Tailoring the properties of correlated oxides is accomplished by chemical doping, pressure, temperature or magnetic field. 
Photoexcitation is a valid alternative to reach out-of-equilibrium states otherwise inaccessible. 
Here, we quantitatively estimate the coupling between a lattice distortion and the charge-transfer excitation in (La$_2$CuO$_{4+\delta}$). 
We photoinduce a coherent La ion vibration and monitor the response of the optical constants in a broad energy range, providing quantitative information on the electron-phonon matrix element that can be compared to theoretical models. 
We propose the same methodology to probe electron-electron interactions in other materials. 
\end{abstract}

\pacs{Valid PACS appear here}

\maketitle

The non-trivial interplay between low- and high-energy degrees of freedom is one of the most distinctive characteristics of strongly correlated systems~\cite{eskes_anomalous_1991,eskes_spectral_1994,molegraaf_superconductivity-induced_2002,novelli_witnessing_2014,mansart_coupling_2013,conte_disentangling_2012,lorenzana_investigating_2013}. 
An elegant way to investigate this phenomenology is provided by pump-probe experiments. 
In Coherent Fluctuation Spectroscopy (CFS), a pump pulse induces a coherent excitation whose effect on the optical properties is monitored in a broad energy window via femtosecond (fs) reflectivity. 
Direct information on the coupling between the coherent mode and the electronic excitations in the accessible spectral range is obtained. 
Coherent modes that can be studied include charge fluctuations in coherent charge fluctuation spectroscopy (CCFS) and lattice vibrations in coherent lattice fluctuation spectroscopy (CLFS). 
For example, Mansart \textit{et al.}~\cite{mansart_coupling_2013} argue that, in order to obtain information on possible pairing actors, rather than using the traditional approach of changing a bosonic excitation and monitor the effect on the superconducting properties (as in the classical isotope effect~\cite{maxwell_isotope_1950,reynolds_superconductivity_1950,gweon_unusual_2004}), one can make use of the reciprocity principle and study how a perturbed condensate affects bosonic excitations. 
Thus, using CCFS, it was found that superconducting quasiparticles are coupled to a high-energy excitation on the Mott scale (2.6~eV). 
Although these experiments provided evidence for the coupling of the condensate quasiparticles to other excitations, the magnitude of the coupling has not been evaluated so far. 

Since the most common mechanism to create coherent excitations involves Raman matrix elements~\cite{merlin_generating_1997,stevens_coherent_2002,zhao_magnon_2004,bao_ultrafast_2004,garrett_vacuum_1997,gridnev2008}, one can apply CFS to virtually any excitation which can be triggered by impulsive or displacive stimulated Raman scattering, i.e. any excitation that is Raman active and has a period longer than the available pump pulse duration. 
In impulsive stimulated Raman scattering (ISRS)~\cite{merlin_generating_1997,stevens_coherent_2002}, the pump pulse can be modeled as a time-dependent electric field acting on the solid, ${\bf E}(t) ={\rm Re}[\bm{\mathcal{E}}(t) e^{\imath E_L t/\hbar}]$, where $ E_L /\hbar$ corresponds to the central laser frequency of the pump and $\bm{\mathcal{E}}(t)$ describes the shape of the 45-fs pump pulse used in our experiments. 
In the following, we use $E$ to denote energies in the eV range and $\omega$ to indicate frequencies of coherent oscillations in the THz regime. 
In ISRS, the Raman interaction converts the incoming optical pulse into an impulsive force $F(t)$ acting on the ions. 
Assuming, for simplicity, a single ion moving in the volume $v_i$, the force in transparent media averaged over one period of the excitation is~\cite{merlin_generating_1997,stevens_coherent_2002} 
\begin{equation}
	F(t)=v_i \frac1{16\pi} \bm{\mathcal{E}}(t) \cdot \frac{\partial \bm{\varepsilon}}{\partial z}(E_L) \cdot\bm{\mathcal{E}}^*(t),
	\label{eq:F}
\end{equation}
where $z$ is the ionic coordinate, $\bm{\varepsilon}$  is the dielectric tensor of the material and its derivative is taken at the pump-laser energy $E_L$. 
The quantity $|{\bm e}\cdot\frac{\partial \bm{\varepsilon}}{\partial z}\cdot{\bm e}^*|^2$ (${\bm e}$ being a unit polarization vector) determines the cross section for spontaneous Raman scattering~\cite{cardona_light_1982,Knoll1995} and its absolute magnitude has been determined only in few cases~\cite{Cardona1990}, with very demanding continuous wave experiments performed at many different laser frequencies. 
Here, we show that this information can be obtained from pump-probe spectroscopy in a single experiment. 
\begin{figure}[tb]
	\includegraphics[width=0.49\linewidth]{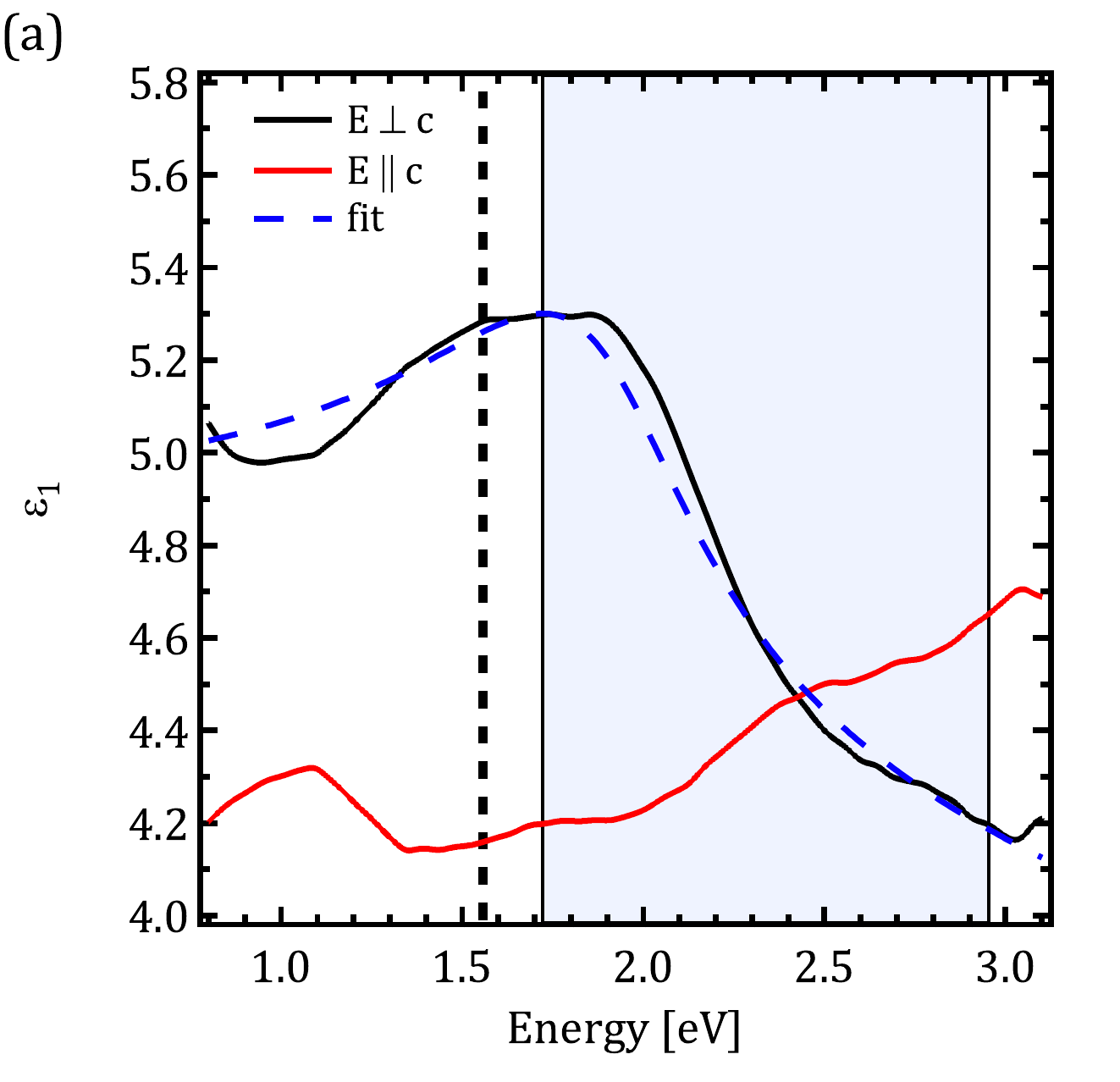}
	\includegraphics[width=0.49\linewidth]{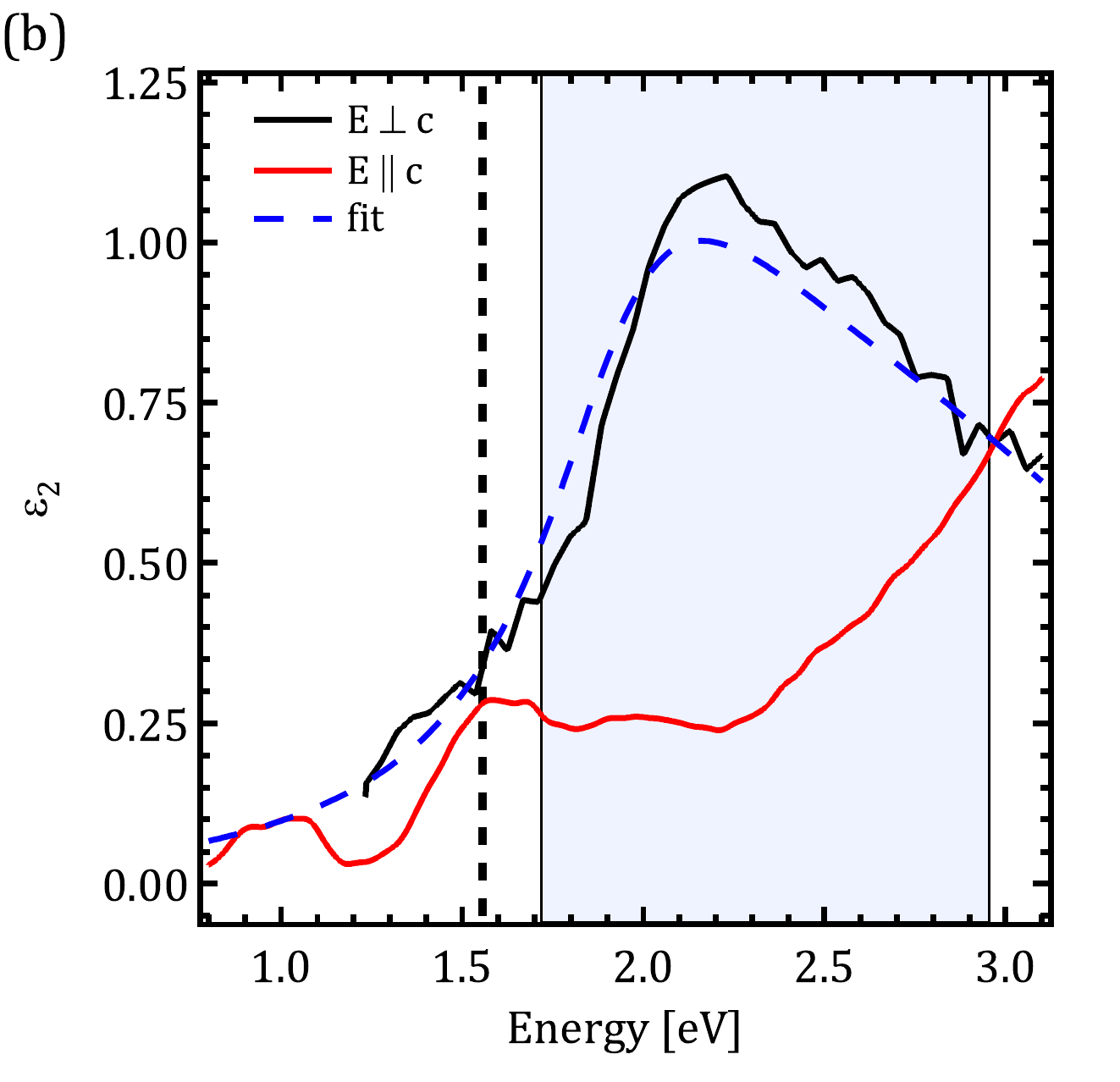}
	\caption{(a) Real part and (b) imaginary part of the dielectric
function $\varepsilon$ from spectroscopic ellipsometry measurements. 
The vertical line indicates the pump energy, the shaded area denotes the probe spectrum. 
The dashed blue lines are fits to the data with the model of Ref.~\cite{falck_charge-transfer_1992}.}
	\label{fig:fig1}
\end{figure}

Assuming a single oscillator of mass $M$ and frequency $\omega_0$, the impulsive force from Eq.~\eqref{eq:F} produces an oscillation with maximum amplitude 
\begin{equation}
	z_M=\frac{v_i {\mathcal F} }{2\omega_0M c}\left|\frac{\partial\varepsilon_{pu}}{\partial z}(E_L)\right|,
  \label{eq:ximi}
\end{equation}
where the fluence ${\mathcal F}$ measures the energy per unit surface arriving at the sample with each pulse. 
The dielectric tensor $\varepsilon_{(pu)}={\bm e}\cdot \bm{\varepsilon}\cdot{\bm e}^*$ in Eqs.~\eqref{eq:F} and~\eqref{eq:ximi} is projected according to the polarization of the pump. 
The ion movement produces oscillations of the reflectivity given by: 
\begin{equation}
	\left(\frac{\Delta R}{R}\right)^{\rm osc}(t,E)=\frac{\partial     \log R}{ \partial\varepsilon_{pr}}\cdot\frac{\partial     \varepsilon_{pr}}{ \partial z}(E) \cdot z(t),
	 \label{eq:ref}
\end{equation}
where $\varepsilon_{pr}$ is the dielectric tensor projected according to the polarization of the probe and $t$ is the time delay between pump and probe. 
The first factor on the right hand side of Eq.~\eqref{eq:ref} is known from Fresnel's equations. 
Taking polarizations such that $\varepsilon_{pr}=\varepsilon_{pu}$ and evaluating Eq.~\eqref{eq:ref} at the time of the maximum amplitude and at the energy of the pump one can eliminate $z$ from Eqs.~\eqref{eq:ximi} and \eqref{eq:ref} and obtain an equation for the squared magnitude of $\frac{\partial\varepsilon_{pu}}{\partial z}(E_L)$ in terms of experimental quantities. 
Eq.~\eqref{eq:ref} then gives access to the full resonant Raman profile. 
In practice, since relating the reflectivity to $\varepsilon$ requires a Kramers-Kronig (KK) transform, it is more convenient to use a model dielectric function as we show below. 

To test these ideas, we performed ISRS experiments on a cuprate parent compound and we report a strategy to obtain the {\em absolute} coupling between low- and high-energy degrees of freedom using CLFS.  
An $A_{1g}$ La-phonon is coherently excited in the (slightly doped) cuprate parent compound (La$_2$CuO$_{4+\delta}$, or LCO) by using fs pump pulses polarized parallel or perpendicular to the Cu-O planes. The consequent real-time oscillation of the optical reflectivity is monitored in the charge-transfer (CT) region to retrieve the energy and polarization dependence of its strength. 
In this situation, the Raman matrix elements carry microscopic information on the electron-phonon matrix elements~\cite{stevens_coherent_2002}. 
Currently available methods for evaluating the electron-phonon coupling allow averaged information over all lattice vibrations to be obtained~\cite{mansart_evidence_2012,Carbone2008,Perfetti2007}. 
In contrast, we show that CLFS provides information on the electron-phonon coupling of {\em specific} phonons in a correlated system, giving strong anchor points to theory and providing a useful knowledge on the selectivity of the pump energy and polarization in the excitation of a mode.

A slightly doped single crystal of LCO was grown as described in the SI, oriented in a Laue diffractometer, cut along a plane containing the a- and c-axis and polished to optical quality. 
The N\'{e}el temperature was determined to be $T_N=260$~K, which corresponds to a doping hole content of $6\cdot 10^{-3}$ or an oxygen content of $\delta = 3\cdot 10^{-3}$ according to Ref.~\cite{kastner_magnetic_1998}. 
Static optical data between 0.5~eV and 4~eV with a resolution of 40~meV were obtained at room temperature using spectroscopic ellipsometry in a fixed polarizer geometry for 70$^\circ$ and 75$^\circ$ angles of incidence. 
Reflectivity measurements with 1.55~eV pump and white-light (1.7-2.8 eV) probe were performed at a temperature of 10~K and 6~KHz repetition-rate in a closed cycle liquid helium cryostat. 
The overall resolution of the experiment was 45~fs (see SI). 
\begin{figure*}[tb]
	\includegraphics[width=0.25\linewidth]{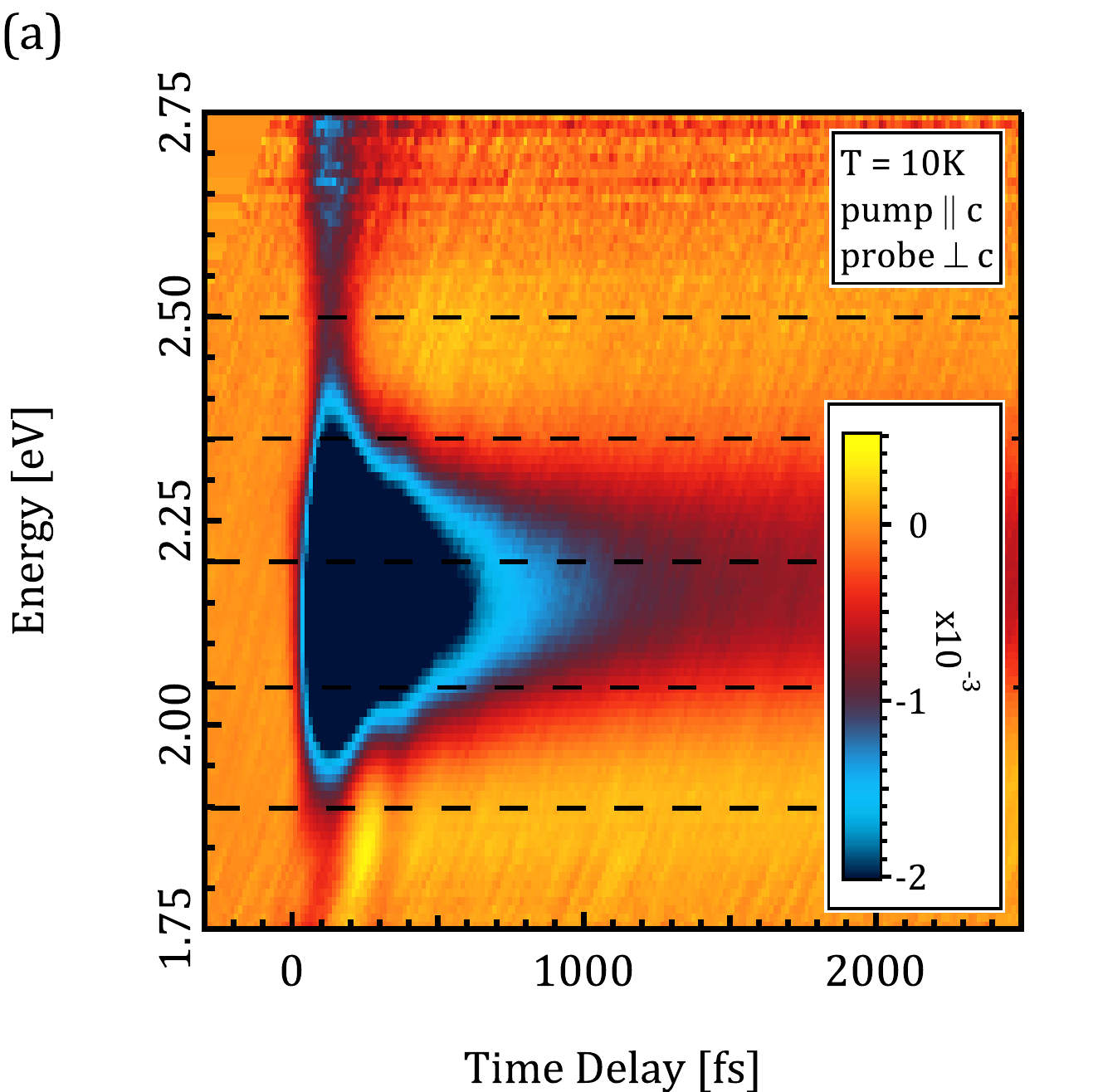}
	\includegraphics[width=0.25\linewidth]{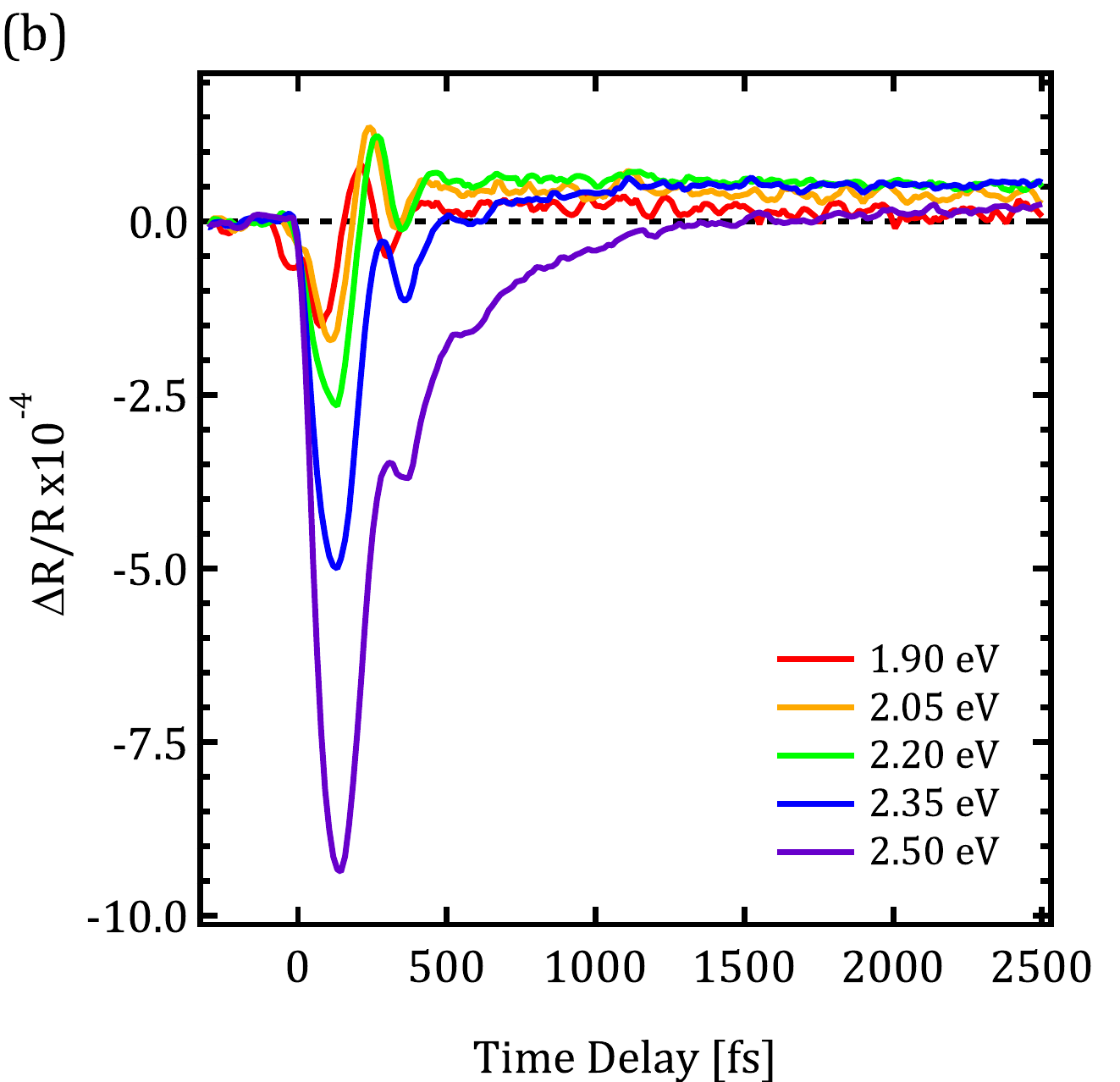}
	\includegraphics[width=0.25\linewidth]{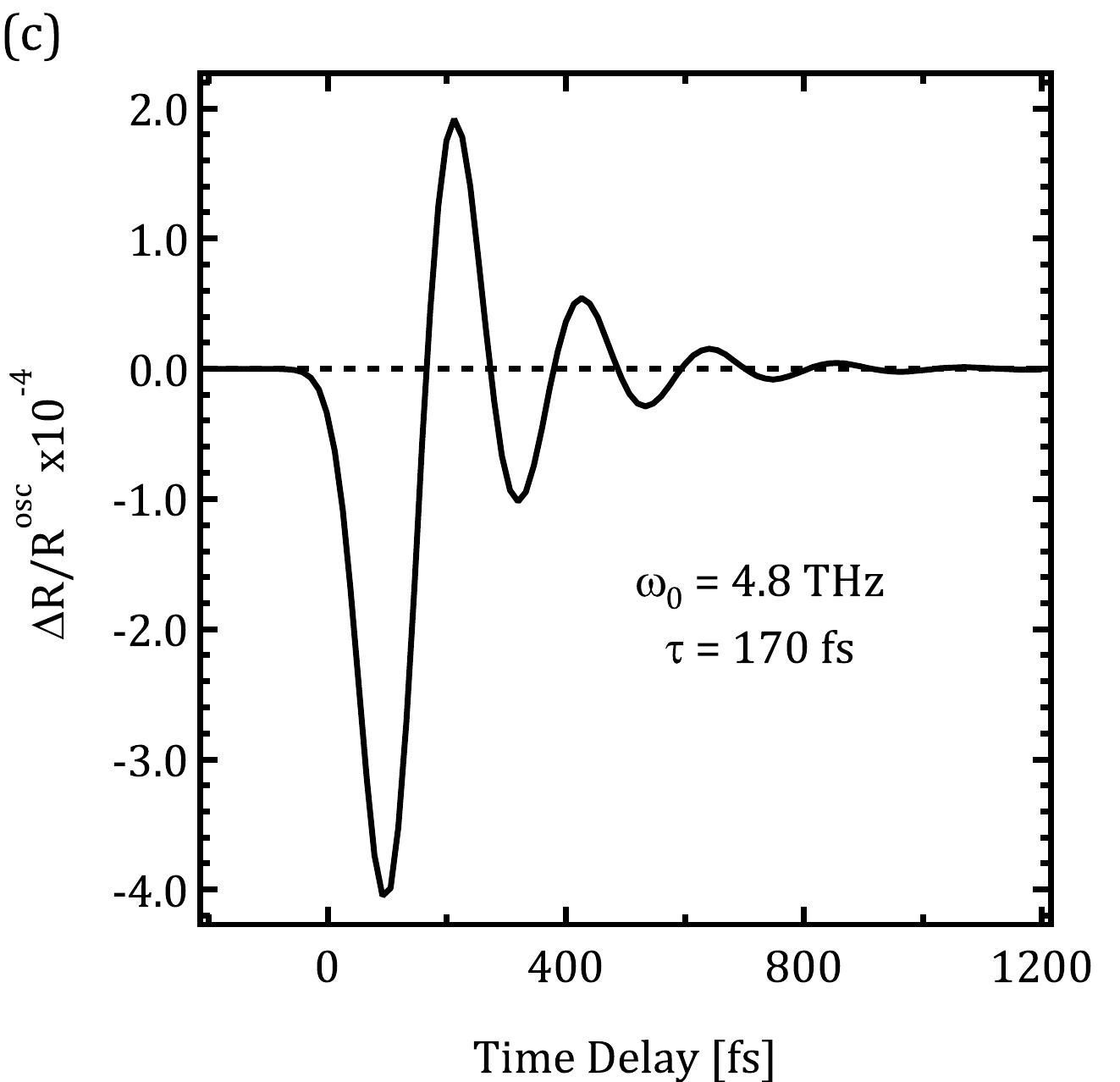}
	\includegraphics[width=0.15\linewidth]{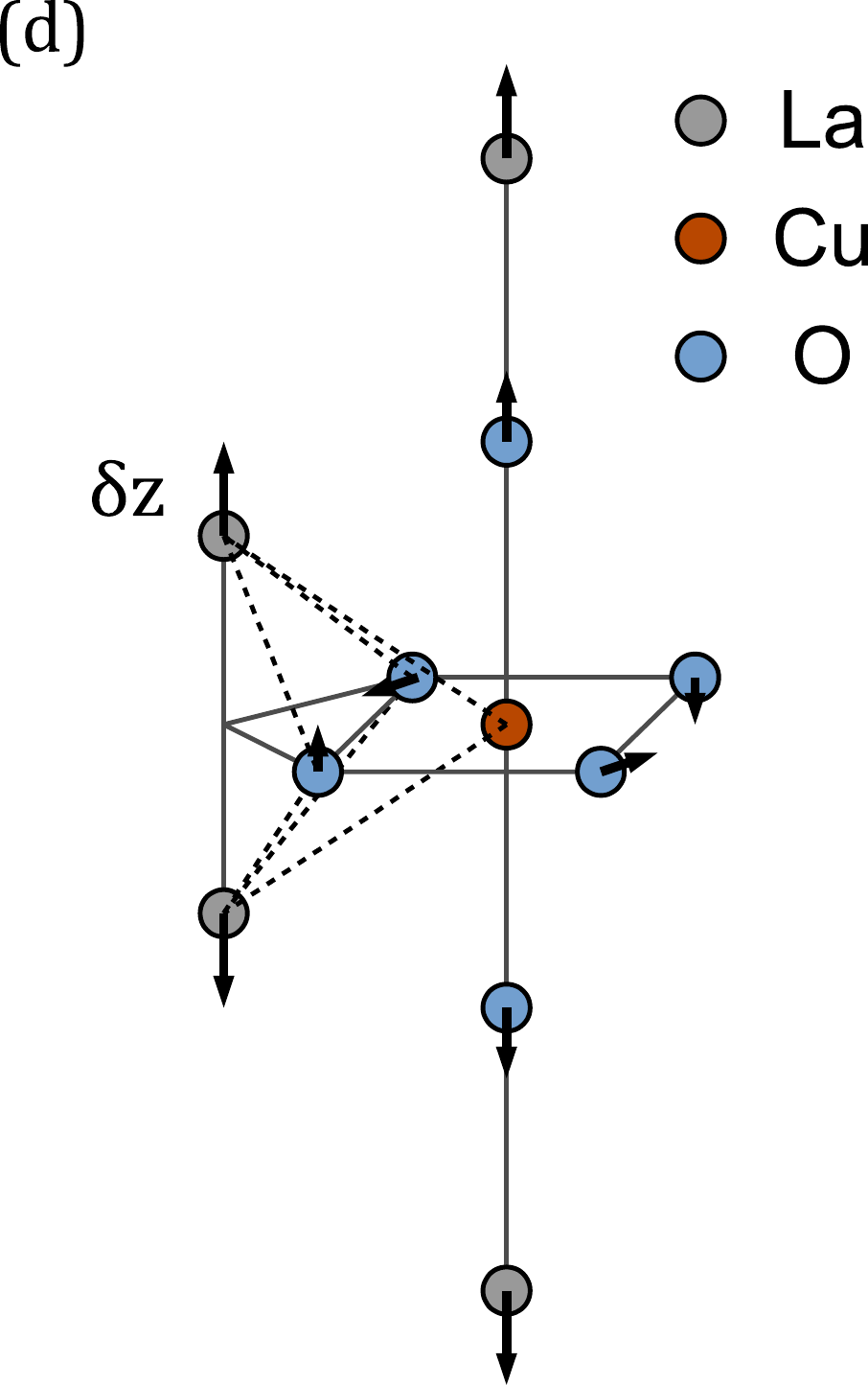}
	\caption{(a) In-plane reflectivity change of La$_2$CuO$_{4+\delta}$, shown in color-coding. 
	(b) Time-dependent reflectivity change for selected probe energies. 
	Traces are integrated over 0.15~eV wide slices centered on the horizontal lines in (a). 
	The pump light is polarized parallel to the c-axis. 
	(c) Oscillatory contribution $\left( \frac{ \Delta R}{R}\right)^{\rm osc}$ of our data for E = 2.1~eV. 
	(d) Eigenvector of the La $A_{1g}$ phonon mode identified as the oscillation in our data, after~\cite{falter_influence_2002}. }
	\label{fig:fig2}
\end{figure*}

Figure \ref{fig:fig1} shows the real and imaginary parts of the dielectric function $\varepsilon$ of the sample under study as measured by spectroscopic ellipsometry, revealing the strong optical anisotropy of the material.
The in-plane imaginary part $\varepsilon_2$ of the dielectric function is dominated by the oxygen $2p$ to a copper $3d_{x^2-y^2}$ CT around 2.15~eV. 
Although not strictly of excitonic nature, Falck \textit{et al.}~\cite{falck_charge-transfer_1992} have described the CT excitation as creating strongly interacting electrons and holes because of the reduced width of the CT peak. 
The finite width of the peak as well as a shoulder around 1.6~eV reveal the presence of excess oxygen in our sample, the actual composition being La$_2$CuO$_{4+\delta}$. 
Comparing $\varepsilon_2$ to static data by Uchida \textit{et al.}~\cite{uchida_optical_1991} we determine that our sample has an effective doping of less than a percent, consistent with the N\'{e}el temperature estimate. 
In contrast to the in-plane behavior, $\varepsilon_2$ is rather featureless along the c-axis. 
As expected, the sample is not a good conductor in this direction, although it shows a feature at the energy of our pump beam. 
To our knowledge, there are no specific transitions assigned to the c-axis spectrum at our pump energy, although the in-plane shoulder present at the same energy can be attributed to a $d$-$d$ transition producing trapped carriers~\cite{ellis_charge-transfer_2008}. 
Nevertheless, it is a general feature of cuprate superconductors that excitation along the c-axis is an effective way of exciting phonons due to enhanced electron-phonon coupling in this configuration \cite{Reedyk1992}. 

For the time-resolved study the sample was excited at 1.55~eV, indicated by the vertical line in Fig.~\ref{fig:fig1}. 
The shaded area highlights the probe range of the experiment. 
We study the case of the pump polarized both in-plane and perpendicular to the plane. 
In the latter case incoherent excitations are suppressed due to the insulating character allowing a better observation of coherent oscillations, so we start by analyzing this case. 
Figure~\ref{fig:fig2}(a) shows the in-plane reflectivity changes of the sample at 10~K for ${\mathcal F} = 7.8$~mJ\,cm$^{-2}$ incident fluence (polarized out-of-plane) as a function of time delay and probe energy for the first 2.5~ps after laser excitation. 

The dominant feature is a strong negative peak around 2.15~eV probe energy that decays within $\sim 0.5$~ps, very similar to the dynamics observed by Novelli \textit{et al.}~\cite{novelli_witnessing_2014} in a comparable experiment. 
The energy maximum coincides with the maximum of $\varepsilon_2$, but the line shape is much narrower, indicating that the CT feature in Fig.~\ref{fig:fig1}(b) is indeed superimposed by the shoulder-like feature around 1.6~eV unrelated to the CT. 
The incoherent part of the response can be understood as being caused by the blocking of excitations across the CT peak due to the depletion of the initial state (population of the final state) by excitations along the c-axis to (from) the intragap states observed in the static data. 
The incoherent peak stretches into a short-lived tail towards higher probe energies (figure top), whereas towards lower energies two vertical lines indicate the presence of a coherent excitation. 
This excitation, which we identify as a coherent La $A_{1g}$ phonon (see SI) and which will be our primary interest in this work, is present over the whole probe spectrum, as can be seen from the time traces for fixed probe energy in Fig. \ref{fig:fig2}(b). 

We used a singular value decomposition (SVD) algorithm to separate the  differential reflectivity matrix into an incoherent part and an oscillatory part $\left(\Delta R/R\right)^{osc}(t,E)$ as detailed in the SI. 
A damped oscillation was fitted to the dominant time traces in the SVD of the differential reflectivity matrix and the weight of each SVD trace was used to reconstruct the oscillatory part of the reflectivity. 
The reconstructed damped oscillation at a fixed probe energy is shown in Fig.~\ref{fig:fig2}(c). 
The decay time is $\tau=170$~fs and the natural frequency is $\omega_0= 4.8$~THz, which agrees with the frequency reported in spontaneous Raman scattering for the A$_1$g phonon in which the La atoms move along the c-axis (see SI), as depicted in Fig.~\ref{fig:fig2}(d). 

\begin{figure}[tb]
	\includegraphics[width=.455\linewidth]{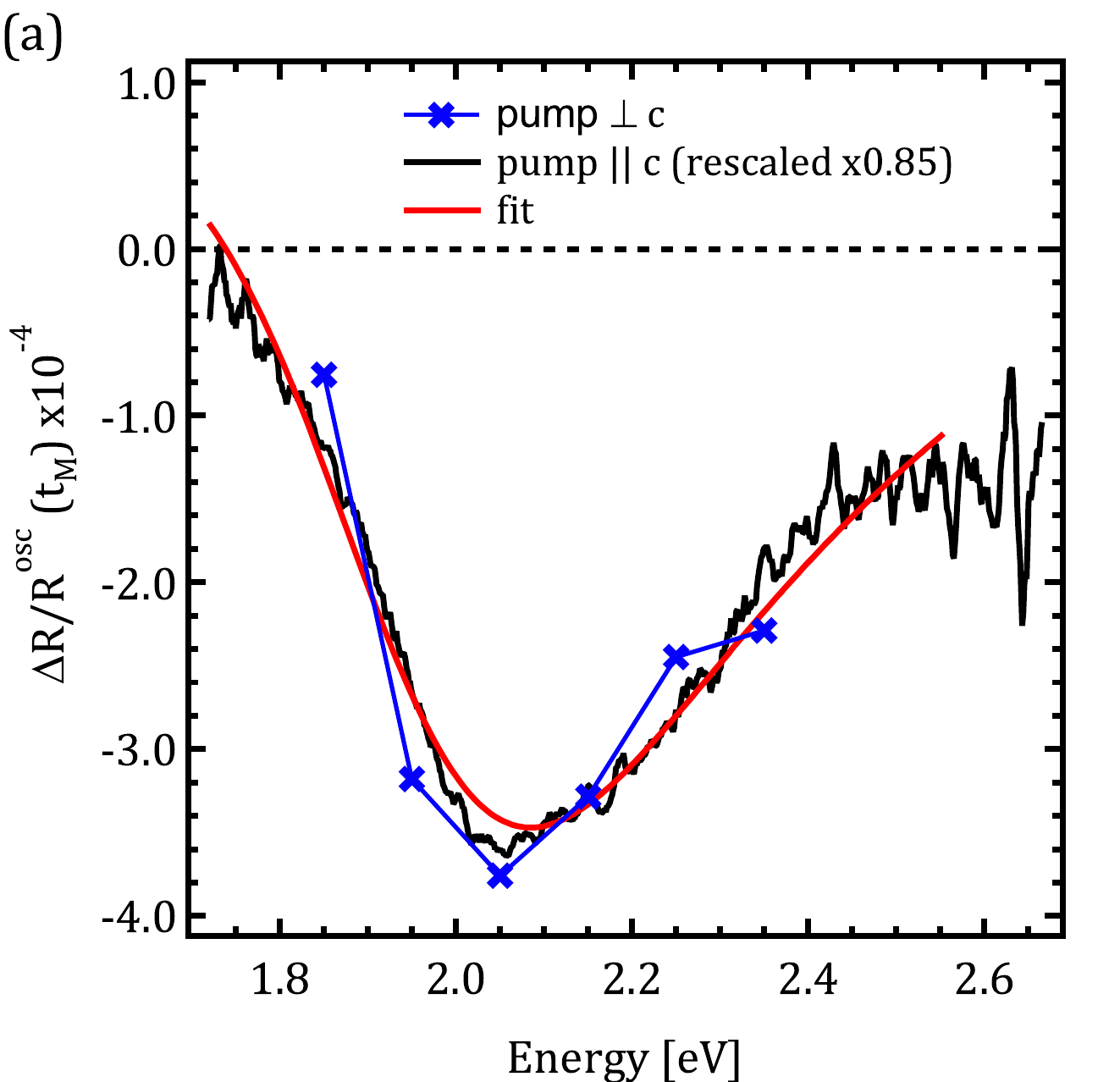}
	\includegraphics[width=.49\linewidth]{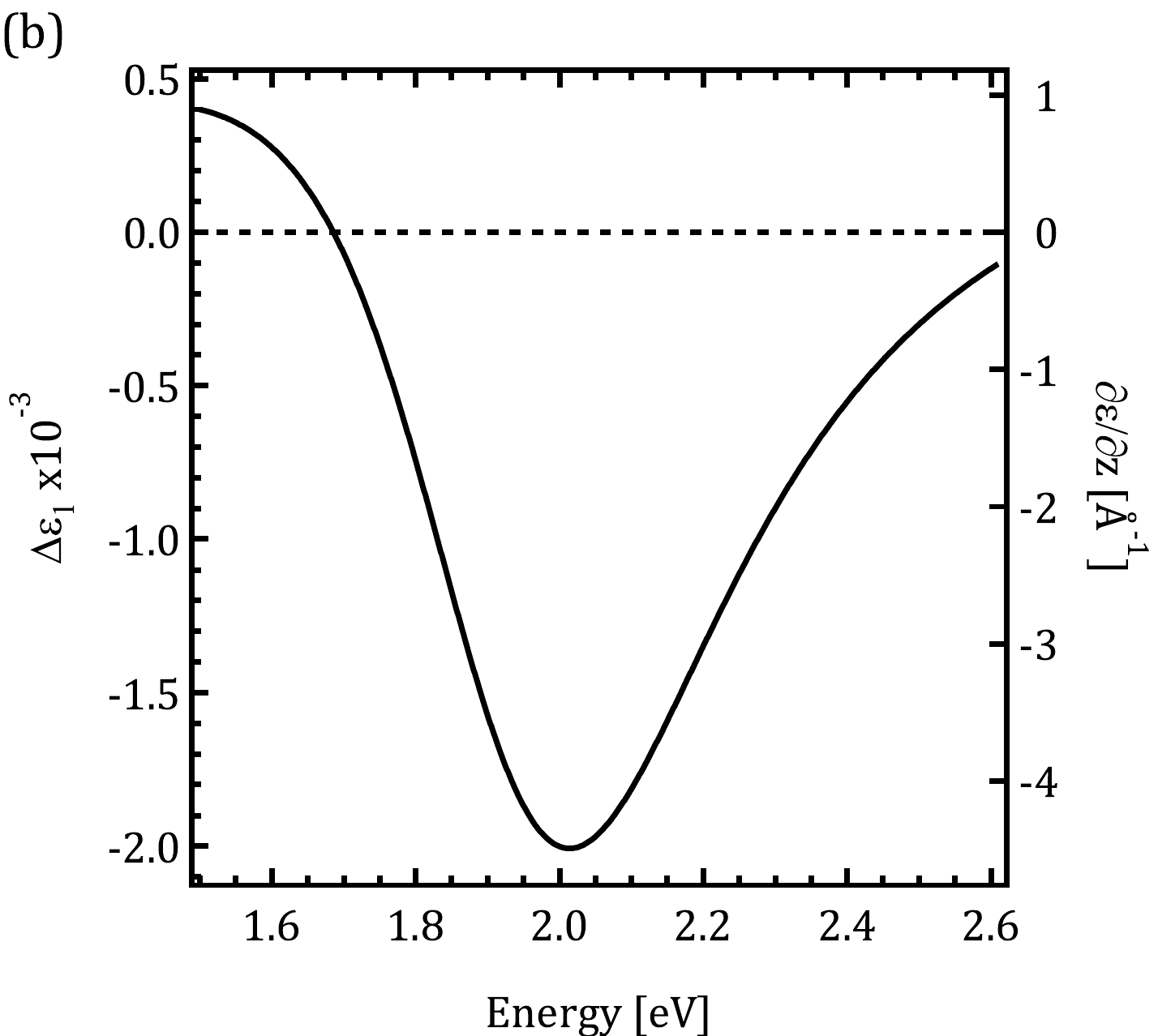}
	\caption{(a) Differential reflectivity energy profile of the coherent oscillation at $t=t_M$ for out-of-plane and in-plane excitation, and fit. 
	The data with the pump $\parallel$ c was rescaled to match the intensity of the data with the pump $\perp$ c.
	(b) Differential dielectric function $\Delta\varepsilon_1$ corresponding to in-plane pump and probe polarizations at $t=t_M$ (left scale). 
	The absolute right scale was obtained as explained in the text. }
	\label{fig:fig3}
\end{figure}

The energy profile of the oscillation extracted from the SVD is shown in Fig. \ref{fig:fig3}(a). 
It shows a resonance around 2.1~eV, which matches the CT feature seen in our static data. 
The energy profile for the same oscillation observed with the pump polarized in-plane is also shown. 
Due to the lower oscillation-to-background ratio, it can only be obtained at a few energy values, integrating over a wide energy range for each point. 
According to Eqs.~\eqref{eq:ximi} and~\eqref{eq:ref}, the absolute Raman profile can be obtained when pump and probe have the same polarization, but the profile shape Eq.~\eqref{eq:ref} depends only on the probe direction. 
Indeed, we obtain very similar line shapes. 
Therefore, we rescale the profile for the pump beam polarized along the c-axis to match the intensity of the case with the pump polarized in-plane, which allows the full differential reflectivity profile to be obtained as if the pump was polarized in the plane. 
The next step is to find the Raman profile $\Delta \varepsilon(E) \equiv \frac{\partial\varepsilon}{\partial z}(E)z_M$. 
While in principle this can be extracted from the differential reflectivity profile using KK transformations, we find it more convenient to extract the information by fitting the data to a simple model. 
As a bonus, this allows the Raman profile to be related to the electron-phonon matrix elements. 

We model the in-plane static dielectric function $\varepsilon$ using an interacting polaron model as described by Falck \textit{et al.}~\cite{falck_charge-transfer_1992}. 
The absorption edge is modeled in terms of transitions between a valence and a conduction band modified by Coulomb interactions which for simplicity are described by a momentum-independent matrix element $U_{eh}$, yielding 
\begin{equation}
	\varepsilon(\omega) = \varepsilon_{\infty} + s \cdot \frac{G(\omega)}{1-U_{eh}G(\omega)}.
	\label{eq:epsilon_model}
\end{equation}
Here, the oscillator strength is pa\-ra\-me\-trized by $s=4\pi e^2 x_{eh}^2 /v_{Cu}$, where $x_{eh}$ is a dipole matrix element between the Wannier orbitals of the two bands and $v_{Cu}$ is the volume per Cu atom. 
We also define the electron-hole Green's function,
\begin{equation}
    G(\omega) \equiv \frac2W \int_{E_g}^{E_g+W}\limits
        d\omega' \left(\frac{1}{\omega+\omega'+i\gamma/2}-
\frac{1}{\omega-\omega'+i\gamma/2}\right),
    \label{eq:osc_strength}
\end{equation}
where, due to the two-dimensionality of the system, we assumed a flat joint density of states $D=2/W$ confined to a band of width $W$ above the gap energy $E_g$. 
The bandwidth $W$ of the absorption band is the sum of the electronic (single-particle) hole and electron bandwidths. 
The linewidth of the transition is given by half the (two-particle) phenomenological damping $\gamma$, which we assume has a simple linear dependence on the frequency, $\gamma=a+b\cdot \omega$. 

Notice that rigorously speaking the model is for a stoichiometric compound, while we have a small oxygen surplus. 
However, the doping is so small that its effect can be phenomenologically absorbed in small changes of the damping and other parameters. 
According to the Born-Oppenheimer approximation it is assumed that the  electronic parameters of the model, like the density of states $D$ and gap $E_g$, depend parametrically on the coherent ionic motion excited by the pump. 

We find that we can simultaneously fit the equilibrium optical response in Fig.~\ref{fig:fig1} and the differential reflectivity profile at the maximum of the oscillation at time $t_M$ in Fig.~\ref{fig:fig3}(a) with only two parameter changes, $\delta E_g = 1.2$~meV, $\delta D = 8.7\cdot 10^{-4}$~eV$^{-1}$. 
Physically, these are also the parameters expected to be most sensitive to the La motion. 
Fig.~\ref{fig:fig3}(b) shows the differential dielectric function $\Delta \varepsilon_1$ corresponding to the oscillation amplitude at $t_M$. 
Neglecting the absorption part, we can represent its value at the pump energy as $\Delta \varepsilon(E_L) = \frac{\partial\varepsilon}{\partial z}(E_L)z_M = 4.4 \times 10^{-4}$. 
This equation together with Eq.~\eqref{eq:ximi} (using for $M$ the mass of one La atom moving in the volume $v_i = v_c/4$, $v_c$ being the orthorhombic unit cell volume) yields a maximum amplitude $z_M=3.2 \times 10^{-4}$~\AA. 
We can use this value to calculate the absolute Raman profile $\frac{\partial\varepsilon}{\partial z} \equiv \Delta \varepsilon(E)/z_M$, shown in the right scale in Fig.~\ref{fig:fig3}(b). 
We are also able to evaluate the electron-phonon matrix elements $\partial E_g/\partial z \equiv \delta E_g/z_M = 2.8$~eV\,\AA$^{-1}$ describing the change of the CT gap with the La motion and $\partial D/\partial z=\delta D/z_M =2.0$~eV$^{-1}$\,\AA$^{-1}$ describing the change in density of states. 

To check if this estimate is reasonable we use a simple ionic model and assume $\partial E_g/\partial z$ is approximately given by the change of the difference in Madelung energy among planar Cu and O atoms when the La is moved in the $z$ direction (see SI). 
We obtain $\partial  E_g/\partial z = 4.3$~eV\,\AA$^{-1}$, in reasonable agreement with the experimental value. 
We attribute the discrepancy to the strong covalency of the Cu-O bond which will make the CT energy less sensitive to the change of Madelung energy with respect to what the simplified ionic estimate suggests. 

To summarize, we have shown that coherent fluctuation spectroscopy can be used to determine absolute Raman profiles much easier than static Raman scattering. 
One potential application of the Raman profile is the manipulation of the electronic ground state by direct excitation of specific phonons~\cite{fausti_light-induced_2011,cavalleri_femtosecond_2001,hilton_enhanced_2007,Stojchevska2014,Yusupov2010}. 
In the case of Raman excitation, the Raman profile allows to find the pump laser energy at which the applied force is maximum, thus allowing to reduce fluence and unwanted heating. 
For example, Fig.~\ref{fig:fig3}(b) together with Eq.~\eqref{eq:F} suggests that exciting at $E_L=2$~eV the oscillations can be enhanced by an order of magnitude using the same fluence $\mathcal F$. 

We also show that the Raman profile can be described via changes in few key-electronic parameters describing the equilibrium optical properties, CT energy and density of states. 
This allows the microscopic electron-phonon matrix elements to be estimated in a correlated system, finding agreement with the upper limit given by a simple ionic model. 
As a bonus, we confirm that ISRS is the mechanism at work on generating the ionic motions. 
We are not aware of previous works where it was checked for a strongly correlated system that the amplitude of the oscillations is correctly given by this mechanism as our analysis implies. 
This is important given that other mechanisms have been proposed in the past~\cite{stevens_coherent_2002}. 
Finally, our results pave the way to a systematic investigation of the interplay between charge \cite{mansart_coupling_2013} and/or structural fluctuations and electronic excitations in complex oxides. 

\begin{acknowledgments}
We acknowledge professor D. van der Marel for allowing us to use his optical spectroscopy equipment for the static experiments. 
Work at LUMES was supported by ERC starting grant USED258697 (F.C.) and the NCCR MUST, a research instrument of the Swiss National Science Foundation (SNSF). 
\end{acknowledgments}

\bibliography{La2CuO4}

\clearpage

\setcounter{page}{1}
\section{Supporting Information}

\subsection{Sample preparation}
Polycrystalline La$_2$CuO$_4$ samples were prepared by a solid state reaction. 
The starting materials La$_2$O$_3$ and CuO with 99.99\% purity were mixed and ground followed by a heat treatment in air at 900-1050$^\circ$C for at least 70 hours with several intermediate grindings. 
The phase purity of the resulting compound was checked with a conventional x-ray diffractometer. 
The resulting powder was hydrostatically pressed into rods (7~mm in diameter) and subsequently sintered at 1150$^\circ$C for 20 hours. 
The crystal growth was carried out using an optical floating zone furnace (FZ-T-10000-H-IV-VP-PC, Crystal System Corp., Japan) with four 300~W halogen lamps as heat sources.
The growing conditions were as follows: The growth rate was 1~mm/h, the feeding and seeding rods were rotated at about 15~rpm in opposite directions to ensure the liquid's homogeneity, and an oxygen and argon mixture at 3~bar pressure was applied during growth. 
The as-grown crystals were post-annealed at 850$^\circ$C in order to release the internal stress and to adjust the oxygen content.

\subsection{Experimental Setup}
Ultrafast reflectivity measurements were performed 
using a Ti:Saph amplified laser system providing sub-50~fs pulses at 1.55~eV with a repetition rate of 6~kHz. 
The reflectivity change of the sample was measured in a pump-probe geometry. 
Of the amplifier output, 0.5~$\mu$J per pulse were used to excite the sample along the c-axis; the temperature was controlled between 10 K and 300 K in a closed cycle cryostat. 
In the probe branch, the beam was delayed in time using a mechanical delay stage and focused into a calcium fluoride (CaF$_2$) crystal, generating a broadband continuum between about 1.7 and 2.9~eV. 
This broadband pulse was focused onto the sample using parabolic mirrors and its reflection collected via optical fiber into a $4f$-spectrometer. 
The latter used a CMOS array synchronized to the incoming laser pulses, enabling us to record the full probe spectrum for every single laser shot. 
Every second pump pulse was blocked by a mechanical chopper, and for each time delay the extracted signal is the relative change in reflectivity, $\Delta R/R$, given by the difference of the probe spectra with and without the pump pulse. For both pump and probe beam, the polarization was controlled using a retardation plate.

The setup has an intrinsic noise level of about 0.1\% RMS. 
The signal-to-noise ratio is improved statistically by repeating each measurement many times as well as binning the signal of several detector pixels, thus averaging in energy. 
In this way, relative reflectivity changes below 10$^{-4}$ can be observed. 
All broadband data presented in this paper have been numerically corrected for the group velocity dispersion experienced by the probe pulse.

\subsection{Singular Value Decomposition Algorithm}
We present a general algorithm to separate differential reflectivity measured as a function of time delay and energy $\left(\frac{\Delta R}{R}\right)(t,E)$ into its different physical components. 
We start by invoking the large separation of energy (or time) scales of the phenomena observed in real time ($\hbar\omega < 0.1$~eV) and the probed window ($\hbar\omega > 1.5$~eV) and in the spirit of the Born–Oppenheimer approximation we assume that the differential reflectivity at high energies can be taken as a parametric function of a set of ``slow'' variables $\xi_\nu(t)$ describing the out-of-equilibrium state produced by the pump. 
Here, $t$ represents the time delay from the arrival of the pump pulse. 
Examples of slow variables $\xi_\nu$ are ionic displacements and ``slow'' charge and magnetic fluctuations, either coherent or incoherent. 
``Fast'' fluctuations produced by the pump are either relaxed after the pump passage ($\sim$ 50 fs) or not resolved. 
We can thus expand the differential reflectivity as a function of the high probe energy $E$ and time delay $t$ as 
\begin{equation}
	\left(\frac{\Delta R}{R}\right)(t,E) = \sum_{i=1}^{N_p}        \left(\frac{\partial \log R}{ \partial\xi_i}\right)(E) \;\; \xi_i(t)
	\label{eq:dR}
\end{equation}
Typically, the sum can be restricted to a small number $N_p$ of processes which contribute significantly. 
In particular, the sum can be partitioned into an oscillatory part $\left( \frac{ \Delta R}{R}\right)^{\rm osc}(t,E)$ and a relaxational part $\left( \frac{ \Delta R}{R}\right)^{\rm rel}(t,E)$. 
 Eq.~\eqref{eq:dR} allows to represent the full two-dimensional data in terms of a few time dependencies of the excitation coordinates $\xi_i(t)$ and their associated energy dependencies $\left(\frac{\partial \log R}{ \partial\xi_i}\right)(E)$ which can be related to the Raman profile for excitation $\xi$ as in Eq.~\eqref{eq:ref}. 
 
In principle one can obtain the decomposition by fitting the experimental data with a model of the form of Eq.~\eqref{eq:dR}, but in practice it is more convenient to use the following algorithm.  
It is natural to consider the data in Fig.~\ref{fig:fig2}(a) as a rectangular matrix $\left(\frac{\Delta R}{R}\right)(t,E)$ of differential reflectivity values with the discrete version of the variables $E$ and $t$ playing the role of indices. 
Singular Value Decomposition (SVD) utilizes the fact that any rectangular matrix $\left(\frac{\Delta R}{R}\right)(t,E) \in \mathbb{R}(m,n)$  (corresponding to data at $m$ time points and $n$ energy values) can be uniquely decomposed into a sum of tensorial vector products~\cite{Trefethen1997} of the form
\begin{equation}
	\left(\frac{\Delta R}{R}\right)(t,E) = \sum_{i=1}^N \lambda_i u_i'(t) \otimes v_i(E) = \sum_{i=1}^N u_i(t) \otimes v_i(E),
	\label{eq:svd}
\end{equation}
The $\lambda_i \in \mathbb{R}$ are the sorted singular values, $\lambda_1 \geq \lambda_{2} \geq \lambda_{3} \geq\ldots$, the $\otimes$ denotes the outer product, $N$ is called the rank of the matrix $\left(\frac{\Delta R}{R}\right)$ and  $u_i(t) \in \mathbb{R}(m) $ and $v_i(E) \in \mathbb{R}(n)$ will be called the canonical time and energy traces, respectively. 
In the second form we have absorbed the singular value in the definition of the canonical time trace $u_i$. 
By construction, the canonical traces are orthogonal, $v_i\cdot v_j = \delta_{ij}$ and $u_i\cdot u_j = \delta_{ij}\lambda_i^2$.  
Restricting the sum over $i$ in Eq.~\eqref{eq:svd}, one finds that only a small number of terms are necessary to reproduce the physically relevant signal in $\left(\frac{\Delta R}{R}\right)$  while the rest contribute to the background noise. 
We can therefore limit the sum to $i=1,..,N_c\ll N$. 
Note that the $u_i(t)$, $v_i(E)$  do not correspond directly to the physical quantities in Eq.~\eqref{eq:dR}. 
This is because the canonical traces are by construction orthogonal while the ``physical traces'' $\left(\frac{\partial \log R}{ \partial\xi_i}\right)(E)$,  $\xi_i(t)$ do not need to be so. 
The latter typically consist of incoherent charge relaxations and/or damped phonon oscillations. 
We thus decompose each of the $N_c$ canonical time traces $u_i(t)$ into $N_p$ physical traces by fitting with a sum of relaxations and damped oscillations representing the physical traces $U_i(t)\propto \xi_i(t)$,
\[
	u_i(t)=\sum_{j=1}^{N_p} a_{ij} U_j(t).
\] 
Inserting this into Eq.~\eqref{eq:svd} we obtain the decomposition of our data in terms of physical traces,
\begin{equation}
	\left(\frac{\Delta R}{R}\right)(t,E) = \sum_{j=1}^{N_p} U_j(t) \otimes V_j(E), 
 	\label{eq:svd2}
 \end{equation}
with $V_j(E)=\sum_{i=1}^{N_c} v_i(E) a_{ij}$. 
We normalize $\sum_{i=1}^{N_c} {a_{ij}}^2=1$ which implies unity norm for the physical energy traces, $V_j\cdot V_j=1$ with $V_j(E)\propto \left(\frac{\partial \log R}{ \partial\xi_i}\right)(E)$.

Eq.~\eqref{eq:svd2} can be separated into an oscillatory part and a relaxation part, according to the character of the model functions $U_j(t)$,
\begin{equation}
	\left(\frac{\Delta R}{R}\right)(t,E) = \left(\frac{\Delta R}{R}\right)^{\rm rel}(t,E)+\left(\frac{\Delta R}{R}\right)^{\rm osc}(t,E).
	\label{eq:svdosc}
\end{equation}
\begin{figure}[tb]
	\includegraphics[width=.49\linewidth]{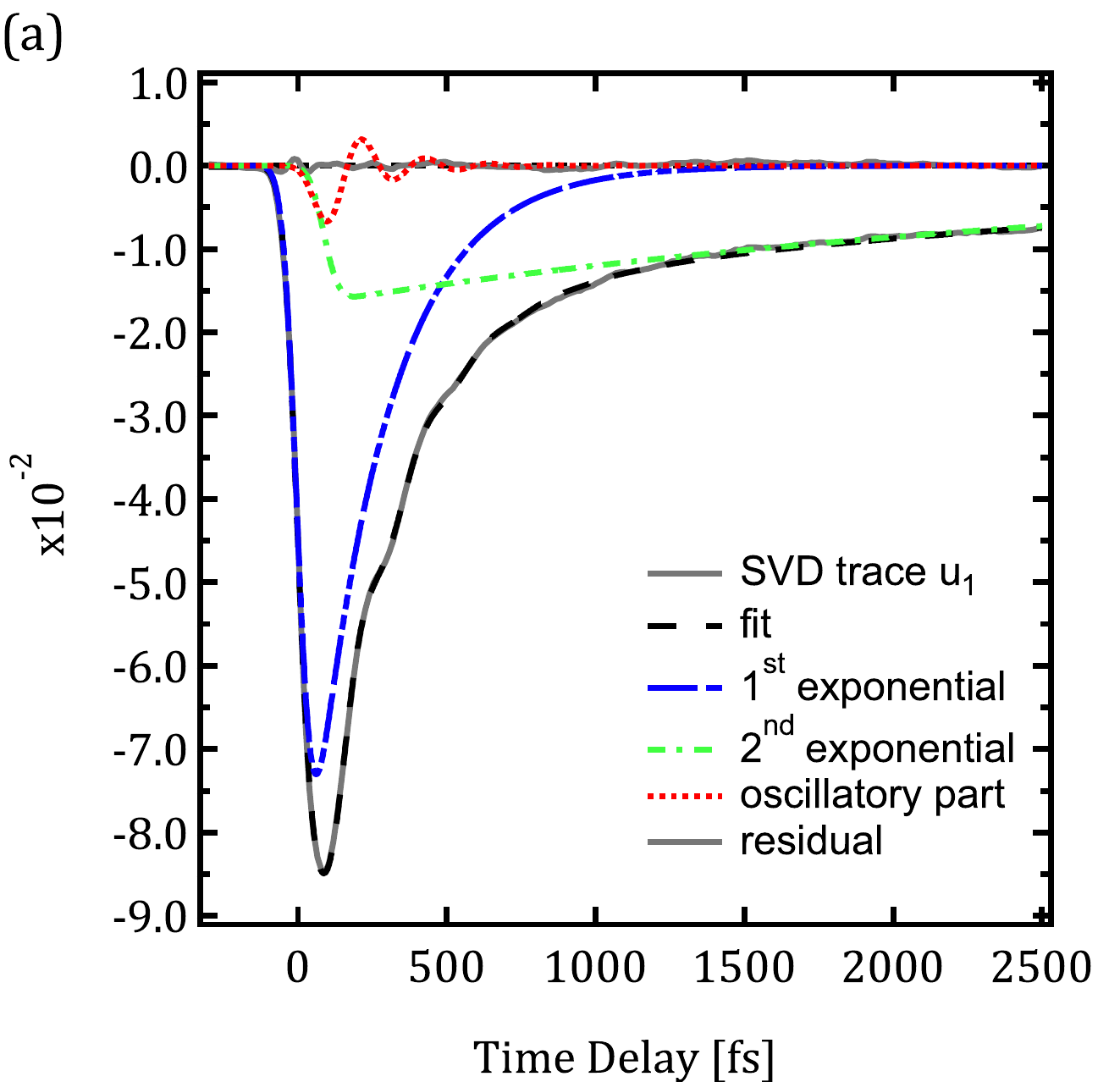}
	\includegraphics[width=.49\linewidth]{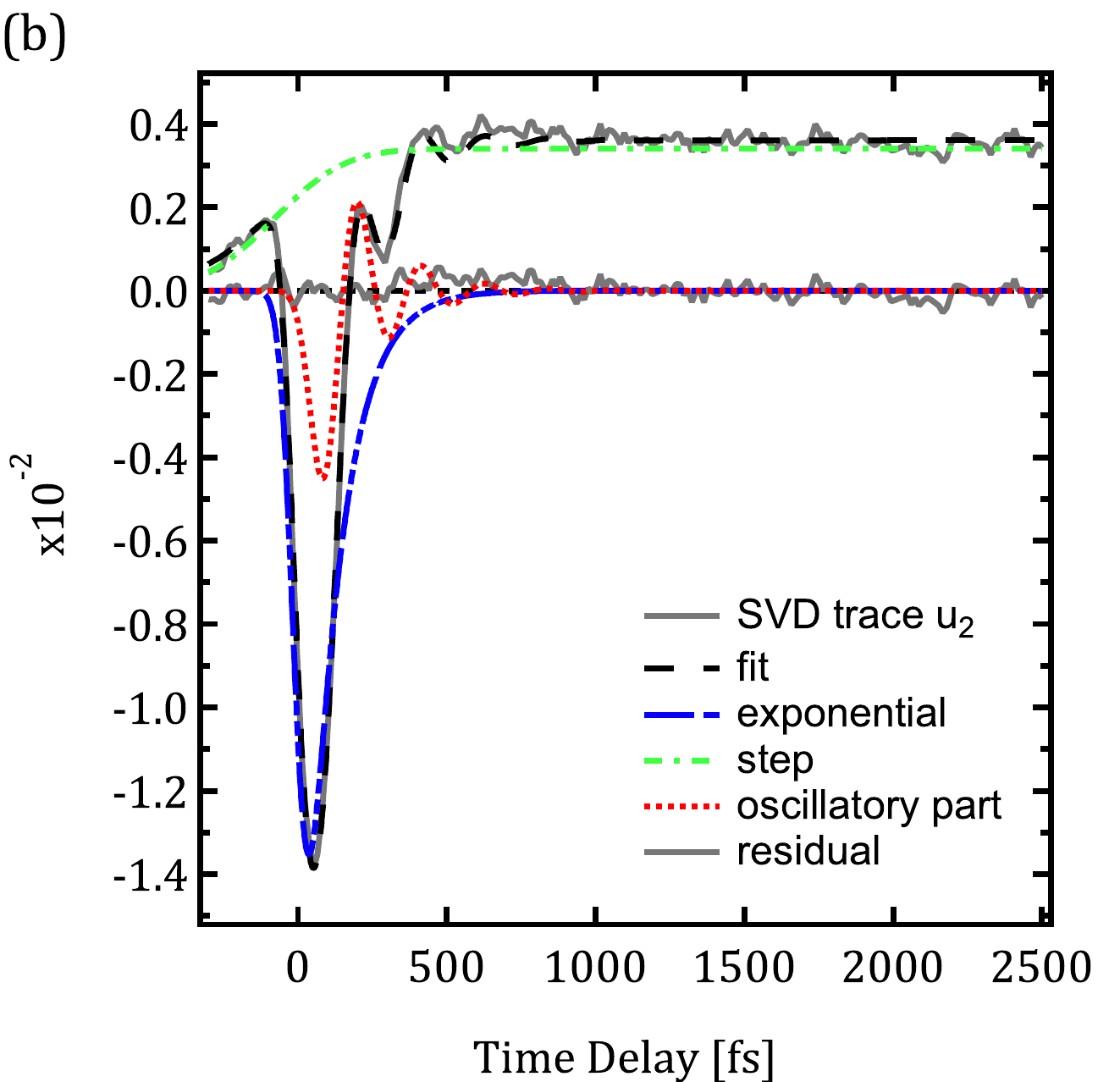}
	\caption{Canonical time traces (a) $u_1(t)$ and (b) $u_2(t)$ obtained from the singular value decomposition. 
	The traces have been fitted with a sum of two exponentials and a damped oscillation.}
	\label{fig:SI_fig1}
\end{figure}
\begin{figure}[tb]
	\includegraphics[width=.49\linewidth]{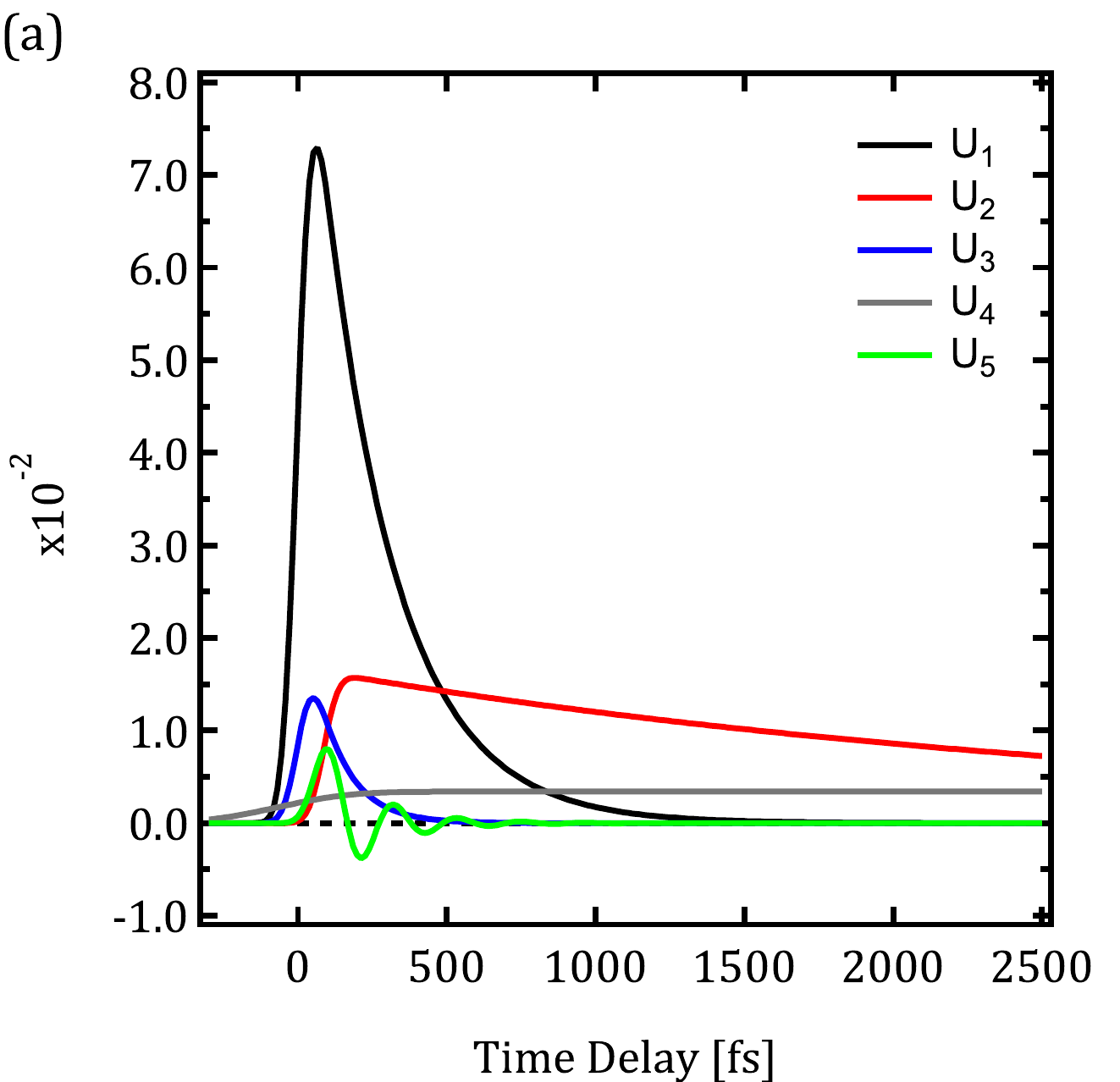}
	\includegraphics[width=.49\linewidth]{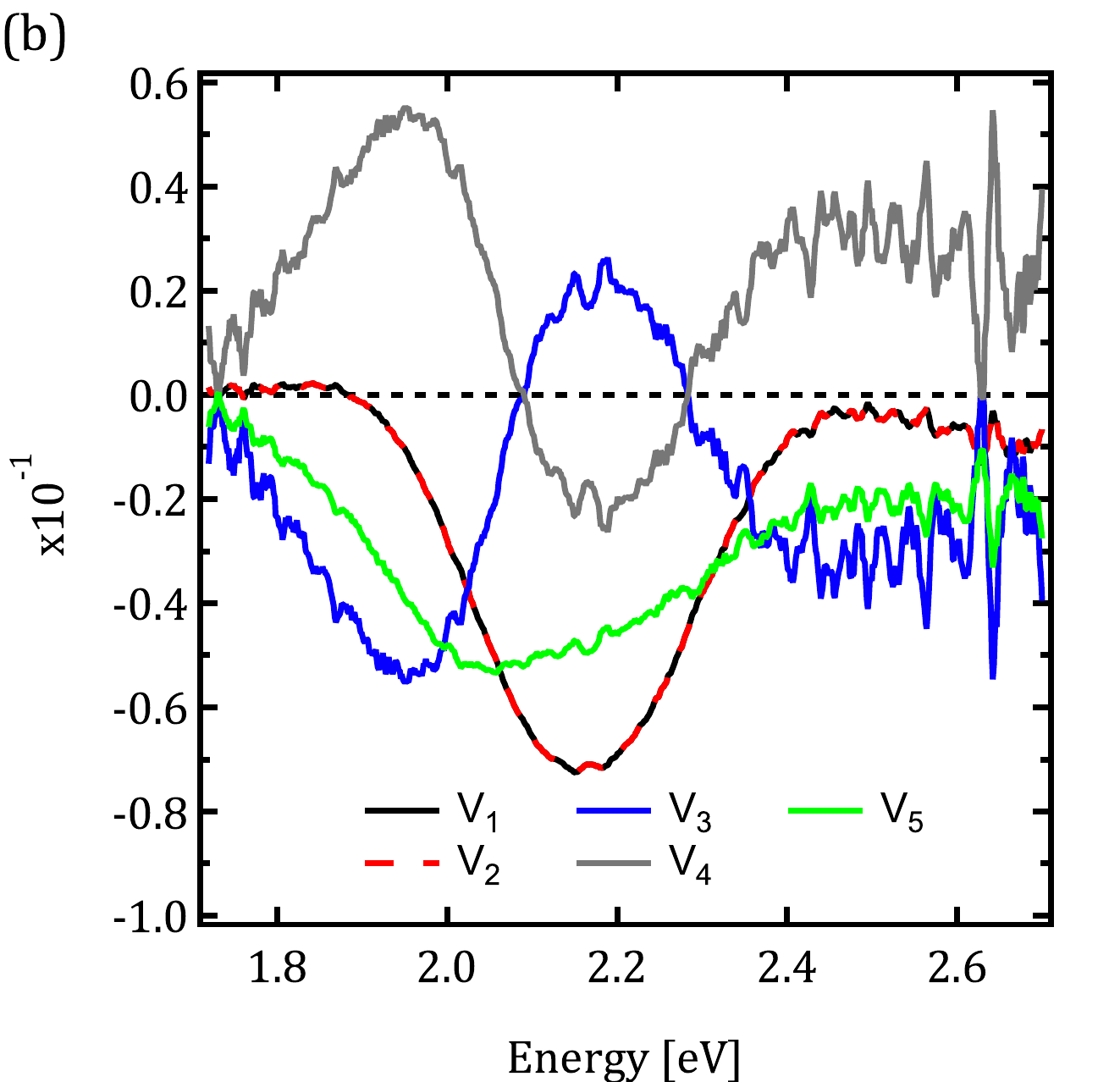}
	\caption{(a),(b) Physical traces $U_i(t)$, $V_i(E)$ extracted from the singular value decomposition. 
	The relaxation is included in $U_1$ to $U_4$, while the oscillation is represented by $U_5$. 
	The energy vectors $V_i$ have been smoothed using a sliding average.}
	\label{fig:SI_fig2}
\end{figure}
\begin{figure*}[tb]
	\includegraphics[width=.32\linewidth]{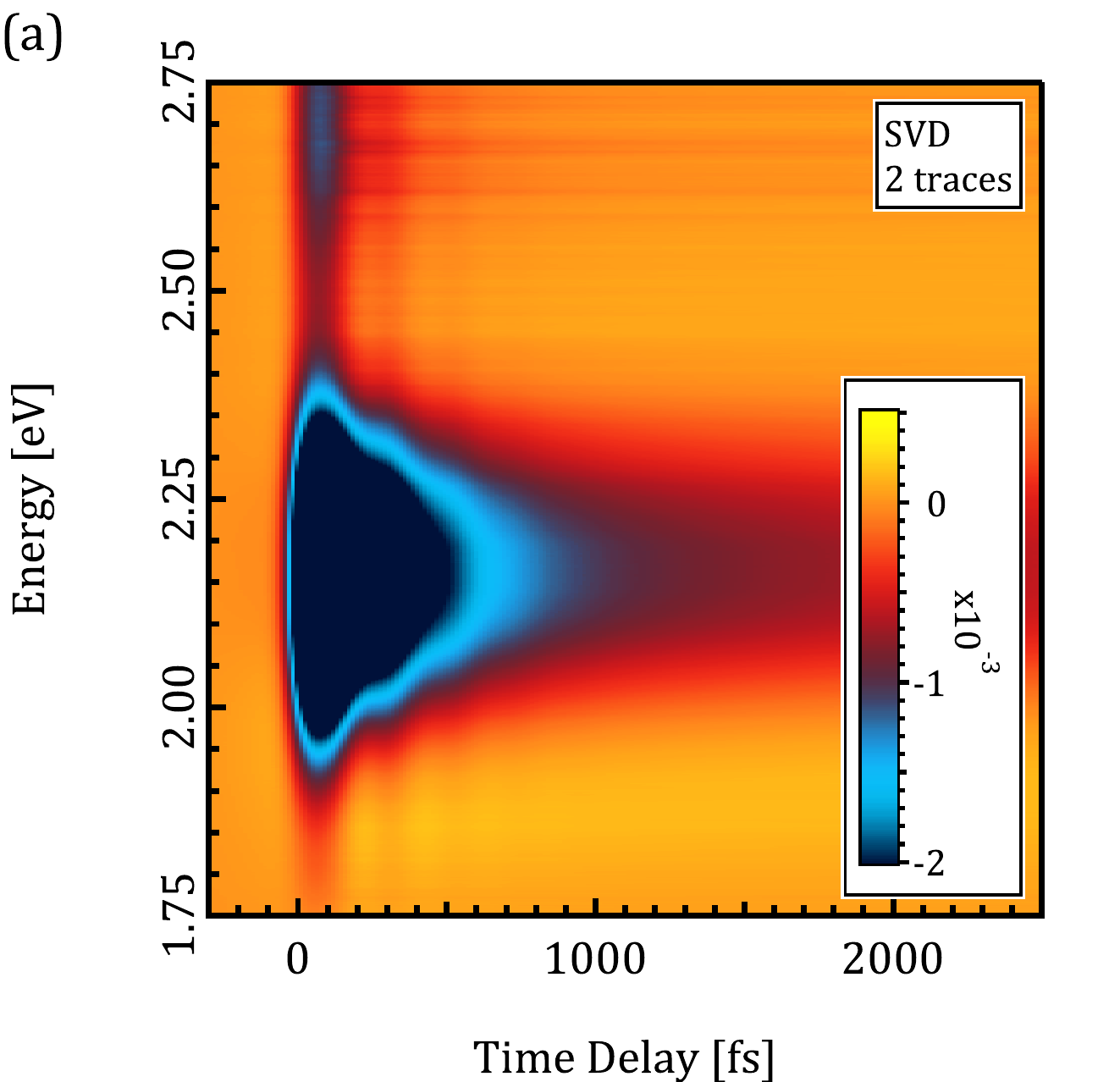}
	\includegraphics[width=.32\linewidth]{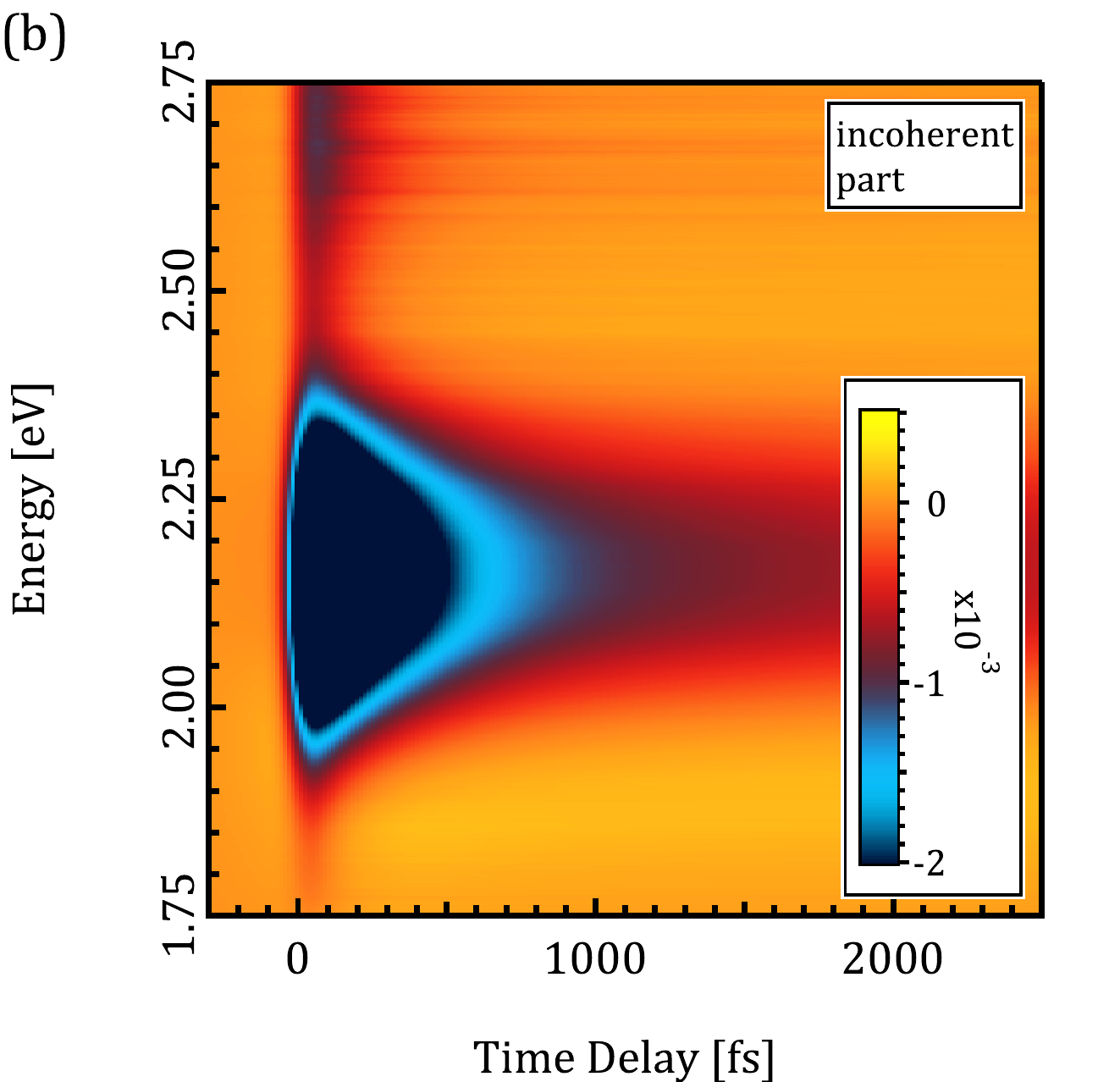}
	\includegraphics[width=.32\linewidth]{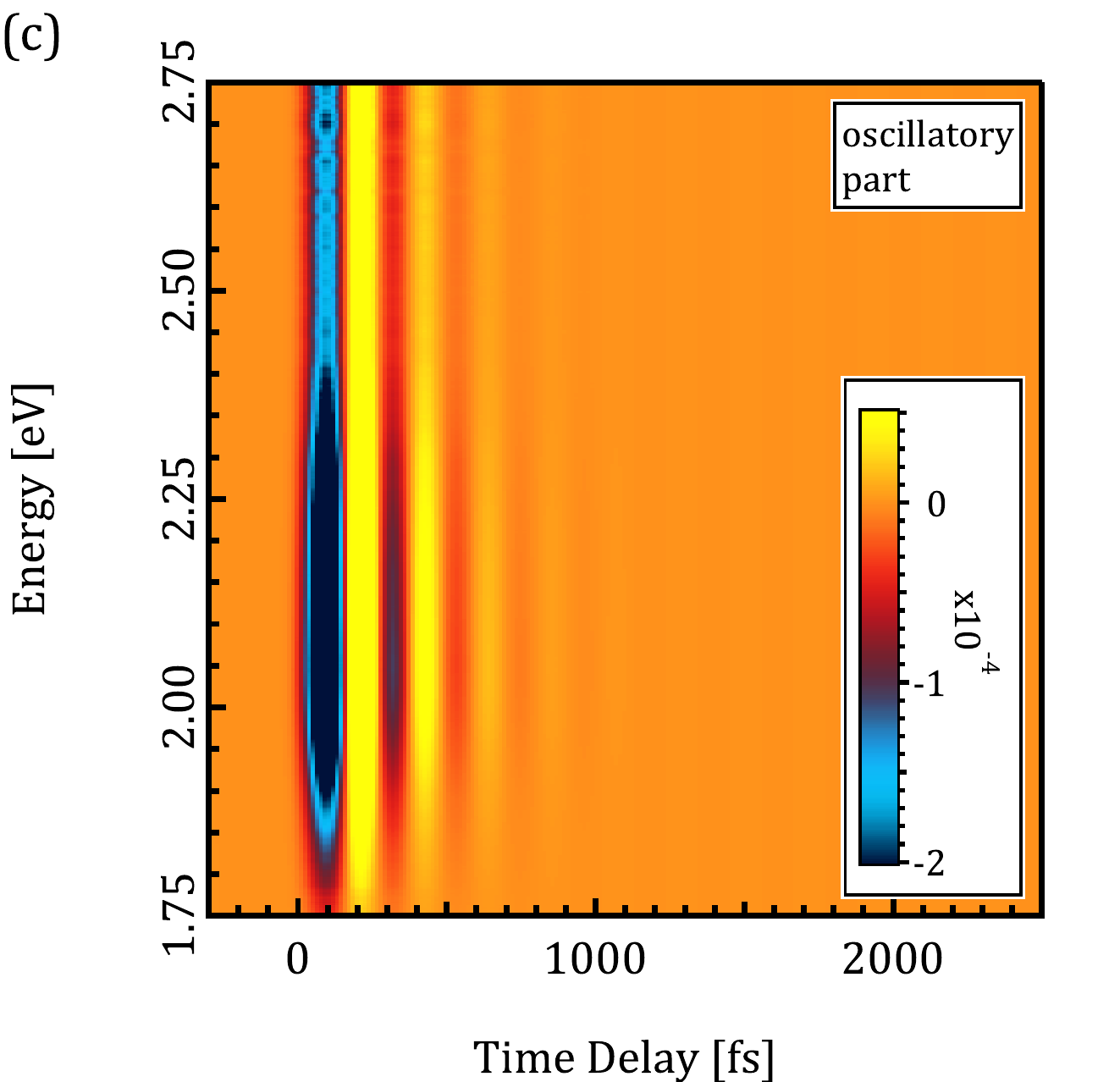}
	\caption{Reconstruction of the data using the SVD results up to second rank: 
	(a) Reconstructed data, compare to Fig.~\ref{fig:fig2}(a). 
	(b), (c) Relaxation and oscillatory contributions according to Eq.~\eqref{eq:svdosc}.} 
	\label{fig:SI_fig3}
\end{figure*}

For the present data we find that the oscillation of interest is present in the first two canonical traces $u_i(t)$. 
Thus, we choose $N_c=2$. 
We find that $u_1$ and $u_2$ can be fitted accurately with the sum of two exponential decays ($u_1$) or an exponential decay and a step funciton ($u_2$), plus one oscillation, as shown in Fig. \ref{fig:SI_fig1}. 
Leaving the frequency of the oscillation to vary independently in the two fits, we find almost the same frequency $\omega=21$ meV (169 cm$^{-1}$) and $\omega=20$ meV (161 cm$^{-1}$) respectively confirming the presence of the same physical trace in the two leading canonical traces. 
Next, we repeat the fit constraining the frequency and the relaxation time to be the same in the oscillatory part so that we can obtain the ``weight'' of the oscillatory physical trace in each of the two canonical traces. 
To get an accurate fit we need to introduce introduce a slight delay (18~fs) of the trace $u_1$ with respect to the trace $u_2$. 
We attribute this is to a numerical error caused by small uncertainties in the correction for group velocity dispersion in the original data. 
Thus, we neglect this small difference in the analysis. 

The physical time traces $U_i(t)$ are shown in Fig. \ref{fig:SI_fig2}(a), where the relaxation is contained in $U_1$ to $U_4$, whereas the oscillation is given by $U_5$. 
Figure \ref{fig:SI_fig2}(b) shows the corresponding energy vectors $V_i(E)$. 
The energy dependence of the oscillation displayed in Fig. \ref{fig:fig3}(a) is determined by the energy vector $V_5$ associated with the oscillation. 
The reconstruction of the data from the SVD for $N_c=2$ and its separation into relaxation and oscillatory components is shown in Fig. \ref{fig:SI_fig3}. 
As mentioned, the main features of the data are reconstructed even when limiting the SVD to second rank. 
Compared to the simpler and more commonly used method of directly using a FFT on the traces of Fig.~\ref{fig:fig2}(b), the SVD is a lot more effective in subtracting the incoherent peak, reduces the noise level, and delivers the energy dependence of the oscillation with ease.

\subsection{Assignment of the coherent oscillation}
The estimates for the oscillation frequency mentioned above allow to identify the  coherent oscillation present in our data as the $A_{1g}$ phonon mode at 156~cm$^{-1}$ present in static Raman data reported by Sugai for the orthorhombic phase of LCO~\cite{sugai_phonon_1989}. 
Previous computations by  Mostoller \textit{et al.} (Table V in~\cite{mostoller_lattice_1990}) show that this mode, listed as 148~cm$^{-1}$ at the $\Gamma$ point in the orthorhombic phase, corresponds to an $A_g$ mode of 135~$cm^{-1}$ (4~THz) at the $X$ point in the tetragonal phase of LCO. 
The eigenvector of the latter was calculated by Falter and Schnetg\"{o}ke (Fig. 7 of~\cite{falter_influence_2002}): 
It is a breathing mode of the La atoms along the c-axis, where the apical O atoms move in phase with the La atoms. 
This mode thus differs from the coherently excited $A_{1g}$ mode in the study of optimally doped La$_{2-x}$Sr$_{x}$CuO$_4$ by Mansart \textit{et al.}~\cite{mansart_coupling_2013}, in which the La and apical O atoms are breathing in antiphase. 
We will not consider the minimal displacements of the in-plane O atoms indicated in~\cite{falter_influence_2002}.

\subsection{Estimate of the electron-phonon matrix element from a point charge approximation}
The ionic displacement induced by the coherent phonon along the c-axis will affect the Madelung energies $\varepsilon_{Cu}$  and $\varepsilon_{O}$ of the copper and oxygen atoms in the ab-plane, changing the charge-transfer energy $\Delta = \varepsilon_{Cu} - \varepsilon_{O} \equiv E_g$~\cite{carbone_real-time_2010}, as well as the density of states $D(\omega$) in the polaron band $W$. 
The electron-phonon matrix element $\frac{\partial E_g}{\partial z}$ reflecting the change in Madelung energies due to the displacement of ions along the c direction can be easily estimated in point charge approximation. 
To this end, we consider the effect of the displacement $\delta z$ of the La atoms closest to the CuO$_4$ plaque on the in-plane CT energy~$\Delta$. 
The atoms in question are the ones sitting above and below the plaque on the positions indicated in Fig.~\ref{fig:fig2}(d). 
The shortest distances to the in-plane atoms are indicated by the dashed lines. 

The energy $\Delta$ is given by difference in Madelung energies of the Cu and O atoms, $\Delta = \varepsilon_{Cu} - \varepsilon_{O} \equiv E_g$~\cite{carbone_real-time_2010}. 
Therefore, $\Delta$ is given as a function of the shortest La distance $z$ from the Cu-O plane by
\begin{eqnarray}
	\Delta(z) &=& \frac{Ze^2}{4\pi\varepsilon_0\varepsilon_R}\cdot \left( \frac{+4}{d_{La-Cu}} + \frac{-8}{d_{La-O}} \right)\notag\\
	&=& \frac{Ze^2}{4\pi\varepsilon_0\varepsilon_R}\cdot \left( \frac{4}{\sqrt{\frac{a^2}{2} + z^2}} - \frac{8}{\sqrt{\frac{a^2}{4} + z^2}} \right),
	\label{eq:delta_z}
\end{eqnarray}
with the La charge $Z=3$ and the unit cell length $a = 3.8$~\AA, which is equivalent to twice the in-plane Cu-O distance. 
Note that there are two La atoms closest to each CuO$_4$ plaque which are nearest neighbors to two Cu atoms and four O atoms each and that the Cu and O atoms carry a single charge, yielding the factors four and eight in Eq. \eqref{eq:delta_z}. 
For the relative dielectric constant $\varepsilon_R$ we use the value $\varepsilon_1(E_{CT})=5.3$ at the CT energy (see Fig.~\ref{fig:fig1}(a)) to account for screening effects. 
Using the equilibrium distance $z_0 = 1.79$~\AA, we find an the el-ph matrix element of $\frac{\partial E_g}{\partial z} \equiv \frac{\partial \Delta}{\partial z} = 4.8$~eV\,\AA$^{-1}$.

\end{document}